\documentclass[
reprint,
superscriptaddress,
amsmath,
amssymb,
aps,
pra
]{revtex4-2}
\usepackage{adjustbox}
\usepackage{makecell}
\usepackage{graphicx}
\usepackage{dcolumn}
\usepackage{bm}
\usepackage{booktabs}
\usepackage{multirow}
\usepackage{amsmath}
\usepackage[colorlinks=true,linkcolor=blue,citecolor=blue]{hyperref}

\begin{document}

\preprint{APS/123-QED}

\title{Noncommutative Index of Measurement-only Entanglement Phase Transitions}

\author{Zhichen Huang}
\affiliation{School of Physics, Beihang University, Beijing 100191, China}

\author{Chunxiao Du}
\affiliation{School of Physics, Beihang University, Beijing 100191, China}

\author{Yang Zhou}
\email[Corresponding author: ]{yangzhou9103@buaa.edu.cn}
\affiliation{School of Physics, Beihang University, Beijing 100191, China}

\author{Zhisong Xiao}
\affiliation{School of Physics, Beihang University, Beijing 100191, China}
\affiliation{School of Instrument Science and Opto-Electronics Engineering, Beijing Information Science and Technology University, Beijing 100192, China}

\date{\today}

\begin{abstract}

Measurement-only models offer an ideal platform for exploring entanglement dynamics in the absence of unitary evolution. Despite extensive numerical evidence for entanglement phase transitions in measurement-only dynamics, the underlying mechanism attributed to non-commutativity among multi-site projective measurements has remained qualitative and coarse-grained. In this work, we identify a quantitative non-commutative index for spatially and temporally homogeneous Pauli stabilizer measurement-only circuits. By applying this index to three representative measurement-only models, we find that the emergence of a volume-law phase is governed by the non-commutative structure of the measurement ensemble, while the transition point is quantitatively determined by the amount of critical non-commutativity. More strikingly, the critical non-commutativity exhibits a linear scaling with the measurement range, independent of the microscopic details of the measurement ensembles. Our findings deepen the understanding of the fundamental mechanism behind the measurement-only entanglement phase transition.
\end{abstract}


\maketitle

\section{\label{sec:Intro}Introduction}
\begin{figure*}[t]
    \centering
    \includegraphics[width=\textwidth]{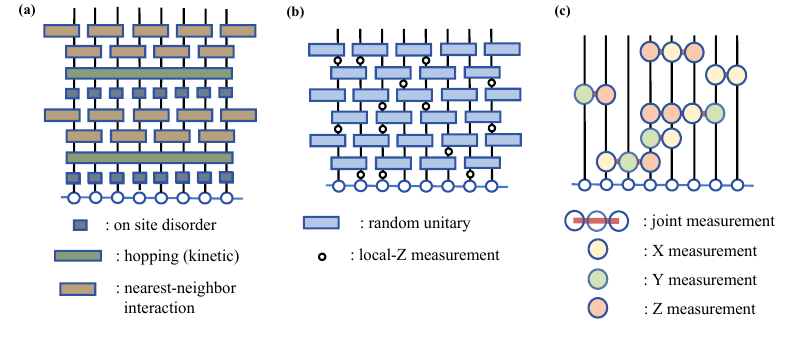}
    \caption{Circuit schematics for unitary, unitary-projective, and measurement-only dynamics. (a) MBL model represented as a quantum circuit, composed of three qualitatively distinct unitary gates corresponding to the three terms in the MBL Hamiltonian. (b) MIPT circuit in a brick-wall structure of random unitary gates interspersed with local-Z measurements applied at random sites. (c) Measurement-only circuit, consisting solely of joint multi-site measurement operations.}
    \label{fig:fig1}
\end{figure*}

Entanglement \cite{PhysRev.47.777, Naturwissenschaften.23, RevModPhys.81.865} is not only a fundamental feature of quantum many-body systems with no classical counterpart, but also serves as a critical resource for various quantum technologies, such as quantum computation, quantum communication, and quantum metrology \cite{NielsenChuang2000, PhysRevLett.70.1895, PhysRevLett.76.722, RevModPhys.91.025001}. As a result, its out-of-equilibrium dynamical behavior in both continuous-time Hamiltonian evolution of conventional condensed matter models and discrete-time evolution of quantum circuit models has attracted increasing interest. The study of entanglement dynamics has led to the discovery of novel entanglement phase transitions (EPTs) \cite{osterloh_scaling_2002, PhysRevX.13.021007}, characterized by singular changes in the entanglement growth rate or in the entanglement properties of steady states. Prominent examples include the many-body localization (MBL) phase transition between a localized phase with logarithmic entanglement growth and a thermalizing phase with algebraic entanglement growth \cite{anderson1958absence, RevModPhys.91.021001, basko_metalinsulator_2006, PhysRevB.75.155111, PhysRevB.82.174411, PhysRevLett.110.260601, PhysRevLett.111.127201, nandkishore2015many}, as well as measurement-induced phase transitions (MIPTs) in monitored quantum dynamics, where measurements interplay with unitary evolution \cite{PhysRevB.99.224307, PhysRevX.9.031009, PhysRevB.100.134306, PhysRevB.98.205136, PhysRevLett.128.010603, PhysRevB.101.104302, PhysRevLett.132.110403, y5r3-tv78, PhysRevLett.127.140601, PhysRevResearch.2.013022, PhysRevB.106.214316, PhysRevB.101.104301, 10.21468/SciPostPhys.14.5.138}. Recently, entanglement dynamics under the combined effects of static disorder and local measurements has also been studied~\cite{LuntPal2020MBL,Sun2025prethermal,Tang2025stronglydisordered}. More intriguingly, measurement alone has the ability to drive a system into different entanglement phases \cite{PhysRevX.11.011030,PhysRevB.102.094204}. The circuit schematics for the three dynamics are shown in Fig.~\ref{fig:fig1}.

In measurement-only systems, time evolution is driven solely by successive multi-site projective measurements without the application of any unitary gates. Despite the absence of unitary evolution, such systems still display several types of phase transition by adjusting measurement parameters, including a sharp transition between area-law and volume-law phases \cite{PhysRevX.11.011030}, which is phenomenologically reminiscent of the MIPT in hybrid circuits; transitions between qualitatively different encoding phases \cite{PhysRevB.102.094204, PhysRevB.108.214302}; and topological transitions \cite{Yu_2025, kuno2022emergence, PhysRevLett.127.235701, PhysRevB.106.104307}. This demonstrates that multi-site joint measurements can both generate and destroy entanglement, in stark contrast to single-site measurements in generic monitored circuits, which act purely as disentangling operations.

For the MBL EPT, it is the competition between disorder and unitary interactions that drives the isolated quantum system to different dynamical entanglement regimes, while the MIPT is generally understood in terms of a competition between the scrambling effects of unitary dynamics and the disentangling effects of single-site measurements. However, the physical origin of entanglement phase transitions in measurement-only systems is fundamentally different from that in unitary Hamiltonian models or unitary-projective circuits. Since no unitary evolution is present, entanglement cannot be attributed to conventional scrambling mechanisms. Instead, it is believed that the key reason for "measurement frustration" that induces scrambling is the non-commutativity among successive multi-site measurement operators. While this picture has been supported by numerical simulations and graph-theoretic arguments, it remains largely qualitative and coarse-grained.

In this work, we systematically investigate spatially and temporally homogeneous one-dimensional measurement-only circuits within stabilizer formalism and introduce a quantitative ensemble-level non-commutativity index. This index measures the probability for two independently sampled Pauli measurements to anti-commute. Numerical calculations are performed for fixed-range factorizable models, correlated XYZ models, and mixed-range factorizable models. Our results reveal that the stationary entanglement phase in measurement-only stabilizer dynamics is controlled by two logically distinct ingredients. First, the emergence of a volume-law entangled steady state is constrained by the overall non-commuting structure of the measurement ensemble. Specifically, when the measurement ensemble can be partitioned into two classes such that operators within the same class mutually commute, the volume-law phase is prohibited and gives way to a critical phase, which in one dimension exhibits logarithmic scaling with system size. This structural constraint can be described by a bipartite frustration graph as discussed in Ref.~\cite{PhysRevX.11.011030}. Second, the location of the phase transition is quantitatively determined by the amount of non-commutativity, as captured by the index introduced in this work. Notably, we observe that the critical non-commutativity grows linearly with the effective measurement range $r$. Taken together, these results establish a clear separation between structural constraints and quantitative control parameters in measurement-only entanglement dynamics and provide a framework for predicting the onset of entanglement phase transitions.

The remainder of this paper is organized as follows. Sec.~\hyperref[sec:levelIcom]{II} introduces the motivation and formal definition of the non-commutativity index $\mathcal{I}(\mathcal{E})$. In Sec.~\hyperref[sec:levelindex]{III}, we validate this index through three representative model classes. Finally, Sec.~\hyperref[sec:Conclusion]{IV} summarizes and discusses our conclusions.

\section{\label{sec:levelIcom}Noncommutative index of random measurement ensembles}

In this section, we construct an ensemble-level index to quantify the degree of non-commutativity in a measurement ensemble consisting of Pauli strings. The measurement ensembles are assumed to be spatially and temporally homogeneous. Spatial homogeneity means that the starting position of a measurement operator is sampled uniformly along the one-dimensional chain, so that the local statistical structure of the ensemble is independent of position. Temporal homogeneity means that the probability distribution of measurement operators is fixed during the evolution. These assumptions remove additional control parameters associated with position-dependent or history-dependent measurement protocols.

In measurement-only dynamics, entanglement generation and rearrangement can arise from sequences of non-commuting projective measurements. While the ordering of measurements along a single trajectory is stochastic, for the homogeneous ensembles considered here, the overall probability for non-commutation is determined by the measurement ensemble itself. This observation suggests that such an index should be defined at the ensemble level and capture the likelihood that two measurements drawn independently from the ensemble fail to commute.

\begin{figure}[t]
    \centering
    \includegraphics[width=0.9\columnwidth]{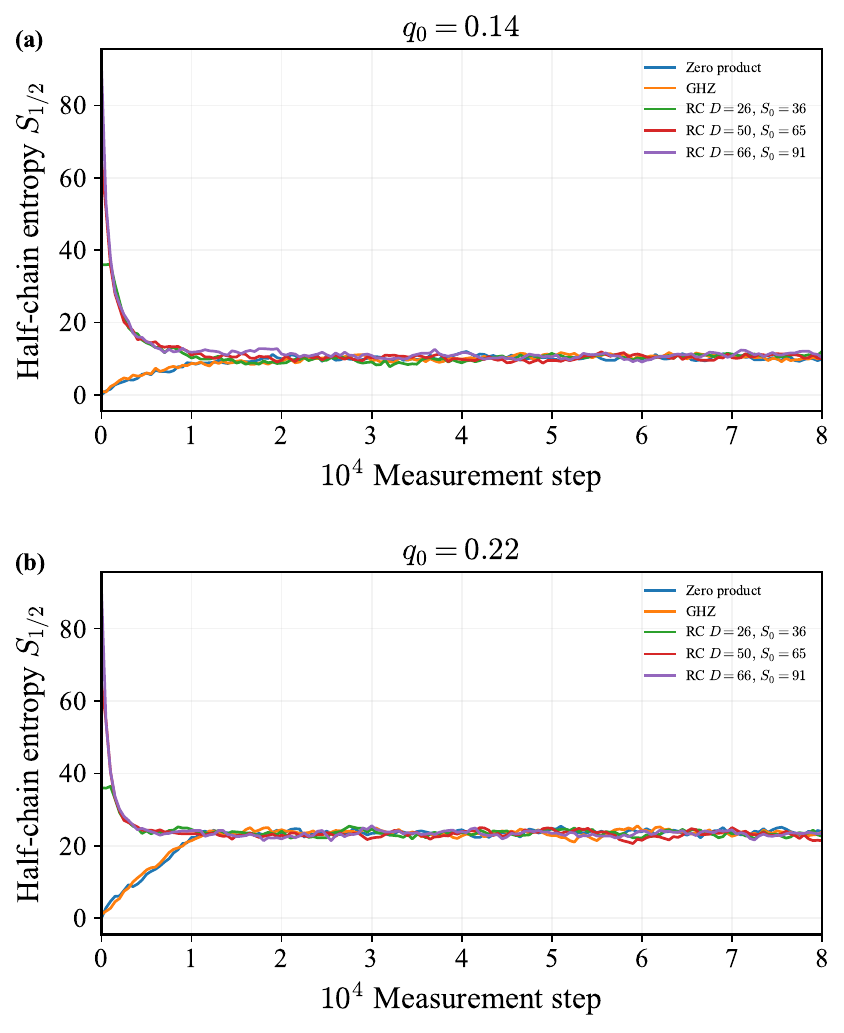}
    \caption{
    Time evolution of the half-chain entanglement entropy $S_{1/2}$ with different initial states. We choose different initial states with different half-chain entanglement entropy and perform random measurements that the Pauli string is randomly drawn from factorizable ensemble with $r=3$. The initial states include pure product state, GHZ state, and random Clifford states with different entanglement entropy $S_0$ which are generated by random Clifford circuit with circuit depth $D$. The total length of qubit chain is $256$, we performed $80000$ times sequential random measurements. All the steady states saturate to a same value of entanglement entropy, implying the irrelevance of initial states for the entanglement phase in measurement-only dynamics.  Here we choose two representative parameter setting of $p_X=p_Y=q_0, p_Z=1-2q_0$ for (a) $q_0=0.14$ (b) $q_0=0.22$.
    }
    \label{fig:fig2}
\end{figure}

To render this idea concrete and computationally tractable, we work within the stabilizer formalism~\cite{PhysRevA.70.052328}, which provides an exact and efficient description of measurement-only dynamics involving Pauli operators. {According to the Gottesman–Knill theorem \cite{gottesman1999group22}, quantum states stabilized by Pauli operators can be classically simulated with polynomial computational cost. More importantly for our purposes, the stabilizer framework makes the role of non-commutativity in entanglement dynamics completely explicit. Within this framework, the effect of a projective measurement is completely determined by whether the measured operator commutes or anti-commutes with the existing stabilizer generators. As a result, non-commutativity directly controls the generation, propagation, and suppression of entanglement. 

A pure stabilizer state of a one-dimensional qubit chain of length $L$ is specified by a set of $L$ independent, mutually commuting Pauli strings
\begin{equation}
    \mathcal{S} =\{g_1,g_2,...,g_L\},
\end{equation}
which generates the stabilizer group $\mathcal{G}=\langle\mathcal{S}\rangle$. The state $|\psi\rangle$ stabilized by $\mathcal{G}$ satisfies
\begin{equation}
    g|\psi\rangle=+1|\psi\rangle \qquad \forall g\in\mathcal{G}.
\end{equation}
Consider a multi-site projective measurement $M$ represented by a Pauli string operator acting on a stabilizer state $|\psi\rangle$. Two distinct cases arise. If $M$ commutes with all generators in $\mathcal{S}$, that is $M\in\langle\mathcal{S}\rangle$, and the state is already an eigenstate of the measurement operator, the measurement leaves the stabilizer group unchanged and has no effect on the entanglement dynamics. In contrast, if $M$ anti-commutes with at least one generator in $\mathcal{S}$, the measurement necessarily modifies the stabilizer group. In this case, one can always choose a generating set such that exactly one generator anti-commutes with $M$. After the measurement, this generator is replaced by $M$, while all commuting generators remain unchanged. 

It should be pointed out that for Pauli measurements on stabilizer states, when the measured operator anti-commutes with the current stabilizer group, the two measurement outcomes occur with equal probability. The post-measurement stabilizer groups associated with the two outcomes differ only by an overall sign of the measured generator, which does not affect entanglement properties. Since the entanglement entropy of a stabilizer state can be calculated from the rank of the corresponding stabilizer tableau, the entanglement dynamics in the present models depend on the commutation structure of the measured Pauli operators rather than on the specific measurement outcomes. Furthermore, owing to the replacement rule described above, the stationary entanglement properties of measurement-only dynamics are independent of the specific choice of initial stabilizers. Under repeated random measurements drawn from a fixed ensemble, initial stabilizer generators that anti-commute with later measurements are progressively replaced by measured Pauli strings chosen from the ensemble. Therefore, after sufficiently long evolution, the stationary entanglement properties are expected to be controlled mainly by the frequency with which measurements drawn from the ensemble anti-commute with the current stabilizer generators rather than by the particular initial stabilizer generating set, as demonstrated in FIG.~\ref{fig:fig2}. This establishes a direct link between entanglement dynamics and the non-commutative structure of the measurement ensemble. Throughout this work, all logarithms are taken to base 2; consequently, the von Neumann entropies and information quantities are measured in bits.

To illustrate this connection explicitly, we simulate the time evolution of the half-chain entanglement entropy along single trajectories, starting from a product state stabilized by $\{Z_1,Z_2,...,Z_L\}$, for a factorizable measurement ensemble which will be further illustrated in Sec.~\hyperref[sec:FE]{III.A}. As shown in FIG.~\ref{fig:fig3}, when the measurement ensemble is fully commuting ($q=0$), the entanglement entropy remains strictly zero throughout the evolution, reflecting the absence of any mechanism for entanglement generation. As non-commuting measurements are introduced ($q>0)$, entanglement growth becomes possible: for small $q$, the entropy fluctuates around a low value and the system relaxes into an area-law steady state. Near the critical point the entropy exhibits slow, approximately logarithmic growth characteristic of a critical phase. For sufficiently large $q$, the entropy grows linearly at early times before saturating, signaling the emergence of a volume-law steady state. While single-trajectory dynamics does not allow one to unambiguously distinguish between critical and volume-law phases in the stationary regime, these examples show that increasing the probability of non-commuting measurement events qualitatively changes the entanglement dynamics. These observations motivate us to define a non-commutativity index that is independent of initial conditions and measurement outcomes, focusing exclusively on the probability that two measurements drawn independently from the ensemble fail to commute. 

\begin{figure}[t] 
\centering 
\includegraphics[width=0.9\columnwidth]{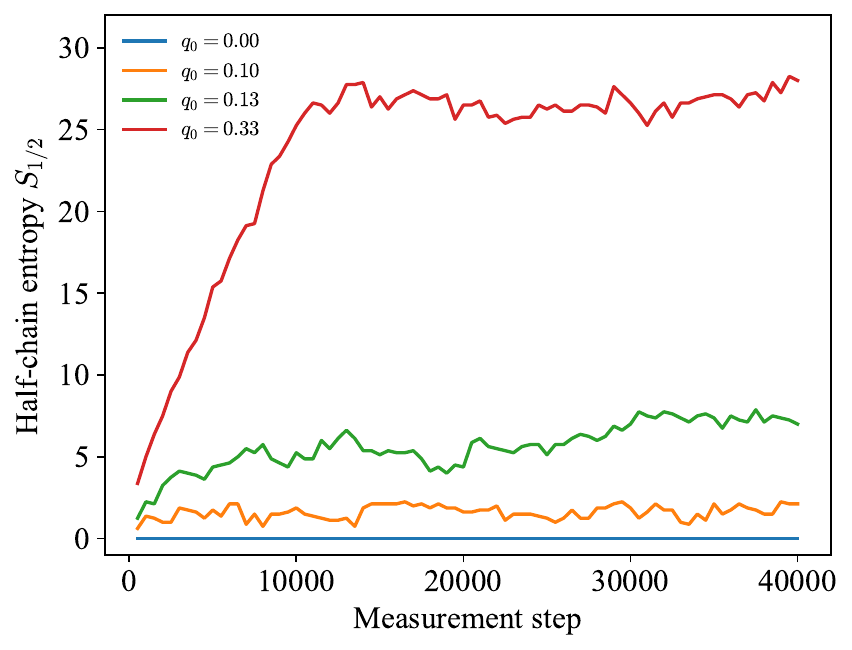} 
\caption{ Time evolution of the half-chain entanglement entropy $S_{1/2}$ along single trajectories for the factorizable measurement ensemble, starting from a pure zero state. Different curves correspond to different values of the probability parameter $q$, with the constraint $p_x=p_y=q$ and $p_z=1-2q$. The blue curve ($q=0$) represents the fully commuting case, where the entanglement entropy remains zero at all times and the steady state is area-law. For $q=0.10$ (orange), non-commuting measurements are present but the system still relaxes to an area-law steady state. At $q=0.13$ (green), close to the critical point, the entropy exhibits a slow growth characteristic of critical behavior. For $q=0.33$ (red), the entropy grows linearly at early times and saturates at a value proportional to the system size, indicating a volume-law phase. } 
\label{fig:fig3} 
\end{figure}

We therefore define the non-commutative index of a measurement ensemble $\mathcal{E}$ as:
\begin{equation}
    \mathcal{I}(\mathcal{E})\equiv\sum_{M_1,M_2\in\mathcal{E}}p(M_1)p(M_2)I(M_1,M_2),
\label{eq:eq3}
\end{equation}
where $M_1$ and $M_2$ are Pauli operators drawn from measurement ensembles $\mathcal{E}$ with probabilities $p(M_1)$ and $p(M_2)$, respectively. The function 
\begin{equation}
    I(M_1,M_2)\equiv\begin{cases}
    0, \qquad[M_1,M_2]=0 \\ 
    1, \qquad\{M_1,M_2\}=0
    \end{cases}
\label{eq:eq4}
\end{equation}
serves as an indicator of anti-commutation. By construction, $\mathcal{I}(\mathcal{E})$ represents the probability that two measurements randomly selected from the ensemble do not commute. It is a purely algebraic property of the ensemble and is independent of measurement outcomes, temporal ordering, and the choice of initial state. Importantly, $\mathcal{I}(\mathcal{E})$ does not depend on the entanglement itself, but instead quantifies the intrinsic capacity of the measurement ensemble to generate frustration through non-commuting measurement events. In the following sections, we show that the critical value of this index governs the entanglement phase transition across a variety of measurement-only models.

\section{\label{sec:levelindex}Non-commutative index for phase transition of measurement only dynamics}

In this section, we examine whether the index defined in Sec.~\ref{sec:levelIcom} can serve as an effective indicator for entanglement phase transitions across three representative classes of measurement-only models: fixed-range factorizable models, correlated XYZ models, and mixed-range factorizable models. All numerical calculations are conducted within the stabilizer formalism. Unless otherwise specified, the systems are initialized in product states stabilized by $\{Z_1,Z_2,\ldots,Z_L\}$, and the entanglement-related quantities are calculated through the rank of the corresponding constrained tableau. 

\subsection{\label{sec:FE}Fixed-range Factorizable Ensembles}

We begin by considering the fixed-range factorizable measurement ensembles proposed in Ref.~\cite{PhysRevX.11.011030}. Here, our goal is not to revisit their phase structure in detail, but to use them as a controlled setting to verify and quantify the role of operator non-commutativity. In a factorizable ensemble, $\mathcal{E}$, each measurement operator is a fixed-range Pauli string, acting on $r$ consecutive qubits without identity operators. The Pauli content on different sites is sampled independently and identically, such that the probability of a given Pauli string factorizes into single-site probabilities,
\begin{equation}
    P_\alpha=\prod_{j=1}^rp_{\sigma_j}, \quad(\sigma_j=X,Y,Z)
\end{equation}
with the normalization constraint $p_X+p_Y+p_Z=1$. As a result, the ensemble is fully specified by two independent parameters, which we choose to be $(p_X,p_Y)$.
Despite their simple construction, factorizable ensembles exhibit rich entanglement dynamics under measurement-only evolution. For sufficiently long measurement range $r\geq 3$, tuning the probability distribution leads to a transition between area-law and volume-law entangled steady states.

For factorizable measurement ensembles with fixed measurement range $r$, the index $\mathcal{I}(\mathcal{E})$ can be evaluated either exactly by exhaustive enumeration of Pauli strings or efficiently via Monte Carlo sampling. In FIG.~\ref{fig:fig4}, we plot $\mathcal{I}(\mathcal{E})$ along two representative directions in parameter space: the symmetric line $p_X=p_Y=q_0$ and $p_Z=1-2q_0$ in FIG.~\ref{fig:fig4}(a), and the margin $p_X=q_0, p_Y=1-q_0$ and $p_Z=0$ in FIG.~\ref{fig:fig4}(b).
It can be seen that, at the same probability setting, longer measurement ranges correspond to larger $\mathcal{I}(\mathcal{E})$. Moreover, along both directions, $\mathcal{I}(\mathcal{E})$ exhibits a single maximum. In case (a), the maximal non-commutativity occurs at the fully symmetric point $p_X=p_Y=p_Z=1/3$, while in case (b) it is reached at $p_X=p_Y=1/2$. These points indicate that, within this factorizable family, ensembles in which non-commuting Pauli operators are most evenly and densely represented have the maximal $\mathcal{I}(\mathcal{E})$.

\begin{figure}[t]
    \centering 
    \includegraphics[width=0.96\columnwidth]{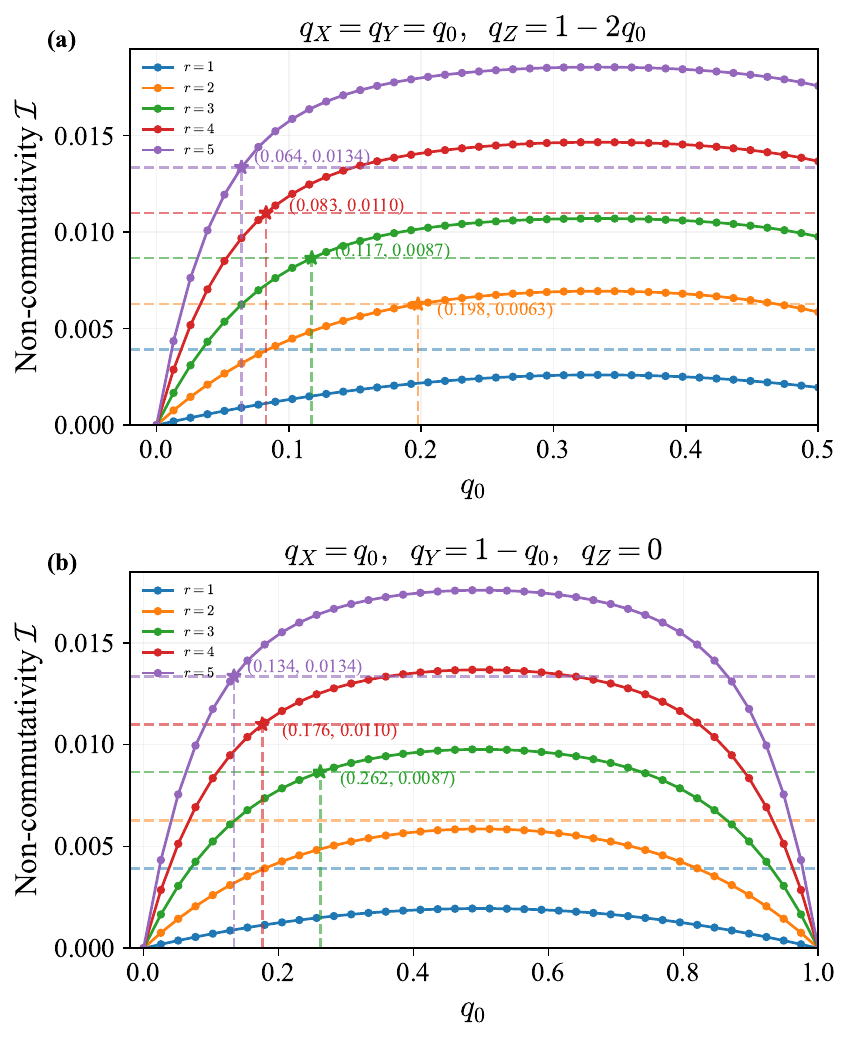}
    \caption{Non-commutativity index $\mathcal{I}(\mathcal{E})$ of the factorizable measurement ensembles with different measurement ranges $r$, plotted along two representative directions in probability space. (a) Symmetric line $p_X=p_Y=q_0$ and $p_Z=1-2q_0$. (b) Marginal line  $p_X=q_0, p_Y=1-q_0$ and $p_Z=0$. In both cases, $\mathcal{I}(\mathcal{E})$ exhibits a single maximum, located at the maximally symmetric points $p_X=p_Y=p_Z=1/3$ in (a) and $p_X=p_Y=1/2, p_Z=0$ in (b), respectively. The horizontal lines indicate the respectively needed $\mathcal{I}$ for phase transition and the corresponding vertical lines depict the phase transition point.}
    \label{fig:fig4}
\end{figure}

The transition points of fixed-range factorizable ensembles were characterized in Ref.~\cite{PhysRevX.11.011030} and can be fitted by
\begin{equation}
r \simeq \frac{k}{\frac{2}{3}-\delta q^2},
\label{eq:eq6}
\end{equation}
where $k=1.16$ and
\begin{equation}
\delta q \equiv \sqrt{\left(\frac{1}{3}-p_X\right)^2+
\left(\frac{1}{3}-p_Y\right)^2+
\left(\frac{1}{3}-p_Z\right)^2}.
\label{deltaq}
\end{equation}
In Fig.~\ref{fig:fig5}, we evaluate $\mathcal{I}(\mathcal{E})$ at the transition point along the symmetric line $p_X=p_Y=q_0$ for different measurement ranges $r$. The resulting critical value $\mathcal{I}_c$ is well described by an approximately linear fit
\begin{equation}
\mathcal{I}_c = k_I r + b_I ,
\label{eq:eq8}
\end{equation}
with $k_I=0.0024$, $b_I=0.0015$, and coefficient of determination $R^2=0.999989$. Part of the linear dependence is naturally associated with the growth of the anti-commutation probability with operator support; the nontrivial observation is that the value extracted at the phase boundary follows a simple linear trend over the accessible range of $r$.

\begin{figure}[t]
    \centering
    \includegraphics[width=0.9\columnwidth]{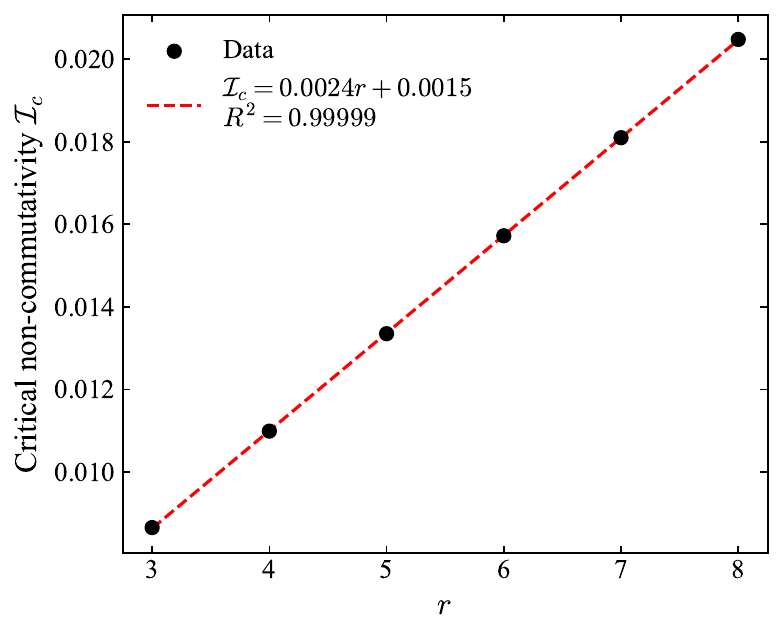}
    \caption{Critical non-commutativity $\mathcal{I}(\mathcal{E})$ of factorizable measurement ensembles evaluated at the entanglement phase transition along the symmetric line $p_X=p_Y=q_0$ for different interaction ranges $r$. The transition points $q_{0,critical}$ are determined from the analytical phase boundary given in Eq.~(\ref{eq:eq6}).}
    \label{fig:fig5}
\end{figure}

While FIG.~\ref{fig:fig5} establishes a clear linear relation between the interaction range $r$ and the critical non-commutativity $\mathcal{I}_c$ along a specific symmetric path in probability space, an important question remains: whether this relation is intrinsic or merely an artifact of the chosen direction. To address this issue, we systematically evaluate $\mathcal{I}_c$ along multiple rays in the measurement-probability simplex. Each ray is defined by an anchor point $(p_X,p_Y,p_Z)\equiv(q_x^{\mathrm{anchor}},\,1-q_x^{\mathrm{anchor}},\,0)$ on the boundary line $p_Z=0$. 

For each anchor, the ray connects the fully symmetric point $(1/3, 1/3, 1/3)$ to the boundary, and the interpolation parameter is fixed such that the distance $\delta q$ matches the critical value $\delta q_c(r)$ determined from Eq.~(\ref{eq:eq6}). This construction allows us to probe the phase transition along inequivalent directions while keeping the transition criterion fixed. As demonstrated in FIG.~\ref{fig:fig6}, all data collapse onto the same straight line as described by Eq.~(\ref{eq:eq8}), showing that the critical non-commutativity depends solely on the measurement range $r$ and is insensitive to the direction in probability space along which the transition is approached.

This is quite striking. Different paths toward the phase boundary in measurement probability space typically represent different operator compositions, so it is not a priori obvious that the relationship between critical non-commutativity and the measurement range should be path independent. Different cuts through the probability simplex yield the same $\mathcal{I}_c(r)$, verifying that $\mathcal{I}_c$ serves as a universal control parameter for the entanglement phase transition in factorizable ensembles independent of microscopic probability distribution $p_X,p_Y,p_Z$ of the measurement ensembles.

\begin{figure}[t]
    \centering
    \includegraphics[width=0.9\columnwidth]{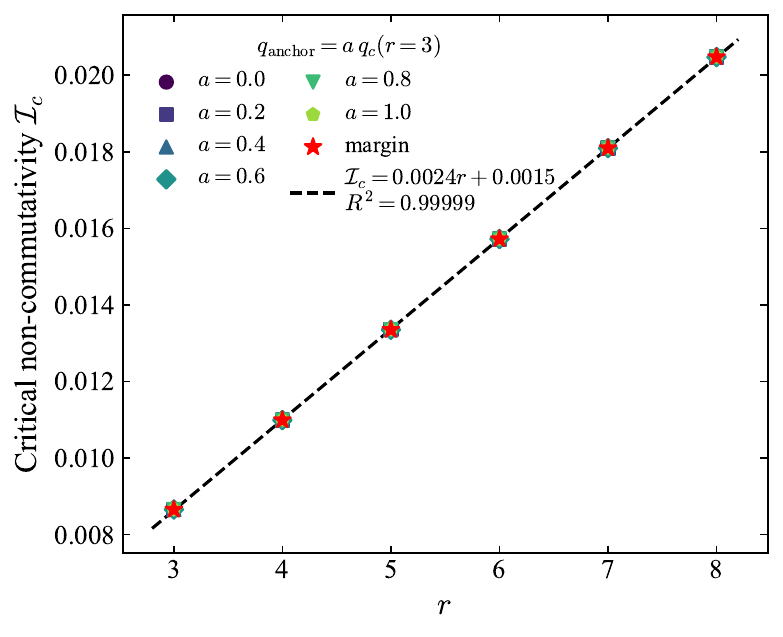}
    \caption{Critical non-commutativity $\mathcal{I}_c$ for factorizable ensembles extracted along different directions in probability space, including one edge of the parameter space and different 6 center-to-boundary lines of the parameter space. All data collapse onto the same linear $\mathcal{I}_c(r)$, same as FIG.~\ref{fig:fig5}.}
    \label{fig:fig6}
\end{figure}

Finally, we mark the entanglement phase transition points and the corresponding critical non-commutativity $\mathcal{I}_c$ along the curves for $r=3$ to $r=5$ in FIG.~\ref{fig:fig4}. The critical non-commutativity $\mathcal{I}_c$ required to induce the phase transition for $r=1$ and $r=2$ is also plotted according to Eq.~(\ref{eq:eq8}). It is indicated that the absence of volume-law phase for $r\leq2$ is mainly because of the lack of non-commutativity. Notably, for $r=2$, the curve intersects the critical line at the point $(q_0,\mathcal{I}_c)=(0.198,0.0063)$ along the symmetric direction $q_X=q_Y=q_0,\ q_Z=1-2q_0$, which suggests the existence of a phase transition. This is consistent with the results of Ref.~\cite{PhysRevX.11.011030}, where a transition from an area-law phase to a critical phase was observed at $\delta q\approx 0.36$, in close agreement with our estimate $\delta q\approx 0.33$. It is suggested that the amount of non-commutativity influences the position of the phase transition, but does not determine whether the non-area-law phase is a volume-law phase or a critical phase. We elaborate this point in the next subsection.

\subsection{\label{sec:XYZ}XYZ Measurement-only Model}

We now turn to the XYZ measurement-only model, which provides a natural and nontrivial extension beyond factorizable ensembles. In this model, each measurement operator acts on $r$ consecutive qubits and is chosen randomly from the set of uniform Pauli strings $X^{\otimes r}$, $Y^{\otimes r}$, and $Z^{\otimes r}$ with probabilities $P_X$, $P_Y$, and $P_Z$. Unlike factorizable ensembles, the Pauli content within a given measurement operator is fully correlated across its support, rendering the ensemble inherently non-factorizable. The XYZ model is therefore useful for testing whether the pairwise non-commutativity index remains informative when the microscopic structure of the measurement ensemble differs from the factorizable case. Furthermore, a key property of the XYZ model is the strong parity dependence of its entanglement dynamics. When the measurement range $r$ is even, a stable volume-law entangled steady state is absent; instead, the system exhibits a critical phase characterized by logarithmic scaling of entanglement entropy with system size. By contrast, for odd values of $r$, a volume-law phase is permitted and the transition from an area-law phase to a volume-law phase is well defined.

\begin{figure}[t]
    \centering
    \includegraphics[width=0.9\columnwidth]{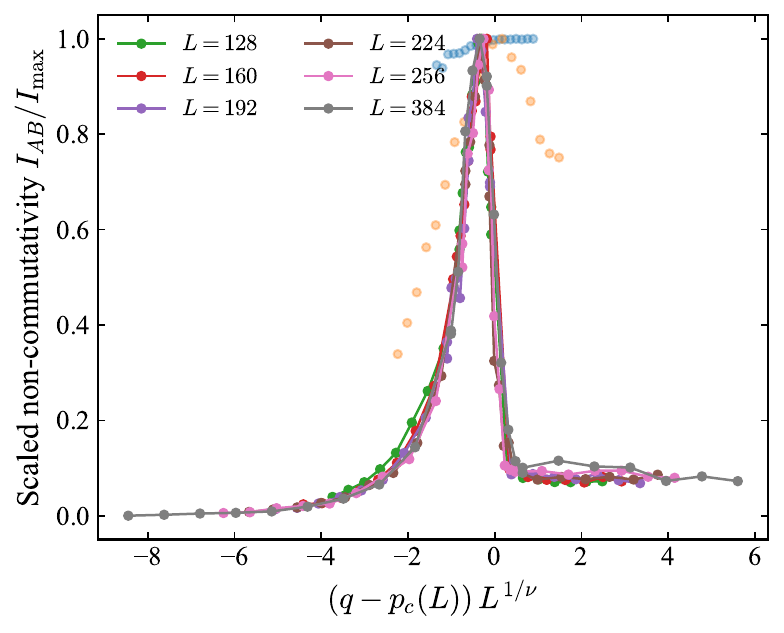}
    \caption{Finite-size scaling collapse of the mutual information $I_{AB}$ for the XYZ measurement-only model at interaction range $r=31$ performed with finite-size shifting of the critical point. The data are rescaled according to (10) with the shifted critical point (11) and $I_AB$ is further normalized by its peak value for each system size $L$, such that all curves attain the same maximum. The optimal collapse yields the correlation-length exponent $\nu=1.3478$ and the thermodynamic critical point $q_c=0.2522$}
    \label{fig:fig7}
\end{figure}

In this work, we mainly focus on how the critical value of ensemble non-commutativity depends on the measurement range. Our calculations are performed at $r=3,7,11,\ldots,31$ for odd $r$ and at $r=2,4,6,\ldots,20$ for even $r$ along the symmetric line $P_X=P_Y=q$, $P_Z=1-2q$. The entanglement transition is characterized by the mutual information $I_{AB}$~\cite{6773024, PhysRevA.56.3470, PhysRevLett.88.017901} between two antipodal regions $A$ and $B$, each of length $L/8$, in a periodic one-dimensional chain of length $L$. The mutual information is defined as
\begin{equation}
    I_{AB}=S_A+S_B-S_{AB}.
    \label{eq:MI}
\end{equation}
Finite-size scaling of the mutual information $I_{AB}$ is employed to determine the entanglement phase transition points in the XYZ measurement-only model, which serves as a sensitive probe of long-range correlations. For a continuous transition, the mutual information is expected to obey the scaling form \cite{PhysRevB.100.134306}:
\begin{equation}
    I_{AB}(L)=F[(q-q_c)L^{1/\nu}],
\end{equation}
where $q$ denotes the tuning parameter along the symmetric line $P_X=P_Y=q, P_Z=1-2q$. $q_c$ is the critical point in the thermodynamic limit, and $\nu$ is the correlation-length exponent.

In finite systems, the apparent transition point exhibits a systematic size dependence. We account for this effect by introducing a finite-size shifted critical point $q_c(L)$, which is assumed to approach its thermodynamic value according to
\begin{equation}
    q_c(L)=q_c(L\rightarrow \infty)+AL^{-b},
\label{eq:eq11}
\end{equation}
where $A,b$ are universal constants for fixed $r$. In practice, we first perform a preliminary collapse of $I_{AB}$ to obtain initial estimates of $q_c$ and $\nu$. Then, the collapse is refined by tuning the shift parameter $A$, adjusting the scaling variable $(q-q_c)L^{1/\nu}$ until the optimal data collapse across different system sizes is achieved. This procedure enables a reliable extraction of the thermodynamic transition point $q_c$ for each measurement range $r$. The corresponding collapsing graph of $I_{AB}$ is shown in FIG.~\ref{fig:fig7}. By numerical calculations, we find that the data are well described by the empirical form:
\begin{equation}
    \delta q_c(r)\simeq A-\frac{B}{r^\alpha},
\end{equation}
with fitting parameters $A=0.203, B=1.924$, and $\alpha=2.219$ for odd $r$ while $A=0.224, B=1.375$, and $\alpha=1.615$ for even $r$.

To illustrate the precision of $q_c(L\rightarrow \infty)$ derived via the above method, we choose $r=31$ as a representative example. Fig.~\ref{fig:fig8} shows the estimated error for $L$ from $128$ to $512$. The statistical uncertainty is estimated by bootstrap resampling over independent stochastic realizations. For this representative case, we obtain
\begin{equation}
    q_c(L\rightarrow\infty)=0.254618\pm0.00030
\label{eq:eq13}
\end{equation}
This result indicates that the extracted transition point is stable within the quoted uncertainty, which demonstrates the robustness of the finite-size critical-point extraction.

\begin{figure}[t]
    \centering
    \includegraphics[width=0.9\columnwidth]{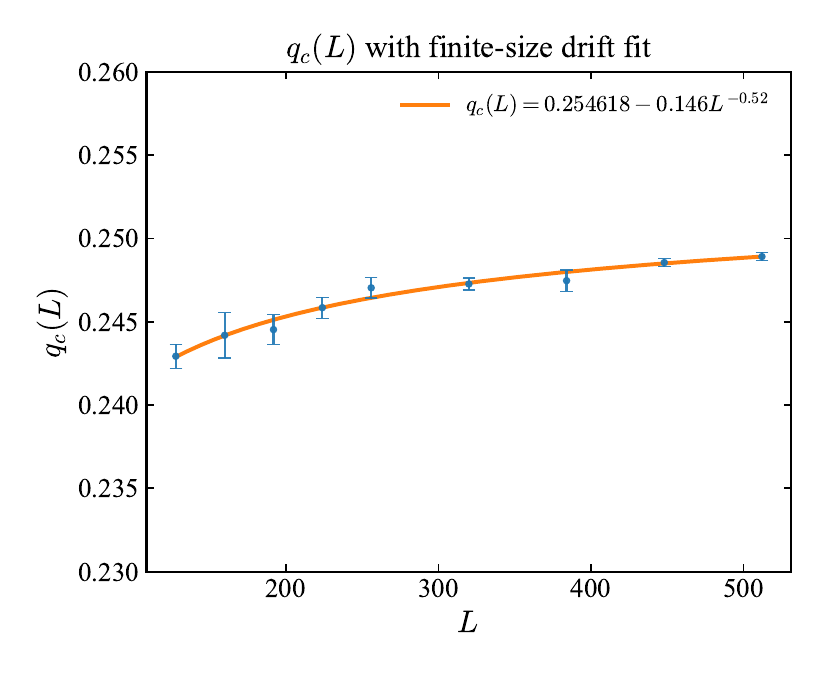}
    \caption{Finite-size drift of the estimated point $q_{c}(L)$ for the XYZ measurement-only model with measurement range $r=31$. The error bars on the data points indicate the bootstrap statistical uncertainty. The solid curve is a fit to Eq.~\eqref{eq:eq11}. This figure is shown as a representative example of the finite-size uncertainty analysis.}
    \label{fig:fig8}
\end{figure}

\begin{figure}[t]
    \centering
    \includegraphics[width=0.9\columnwidth]{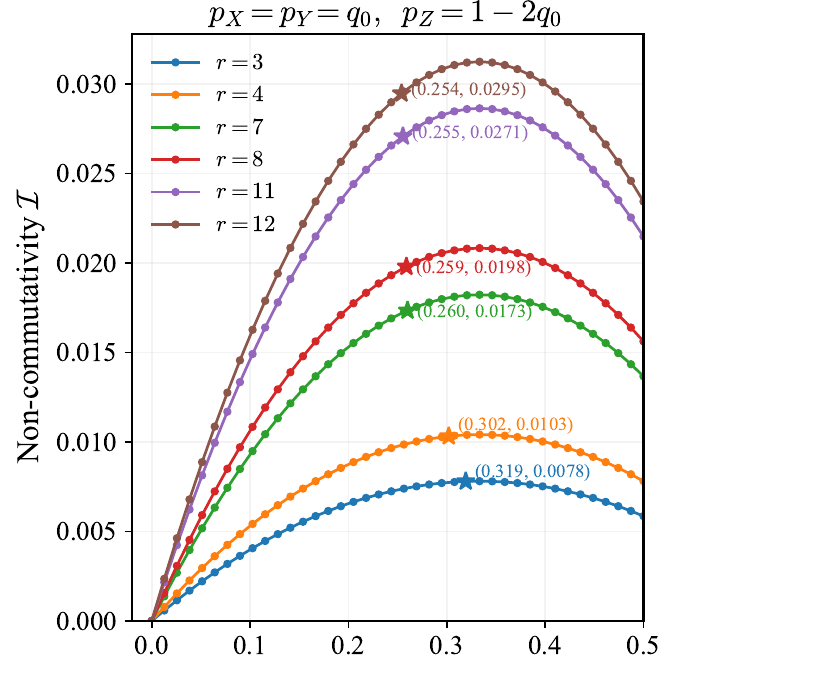}
    \caption{Non-commutativity index $\mathcal{I}(\mathcal{E})$ of XYZ measurement-only models with measurement range $r=3,4,7,8,11,12$, plotted along symmetric directions in probability space. This figure demonstrates that both even and odd $r$ attain enough Non-commutativity for a phase transition.}
    \label{fig:fig9}
\end{figure}

Having determined the critical points $q_c(r)$ of the XYZ model numerically, we plot the $\mathcal{I}$–$q_0$ curves along the symmetric direction $p_X=p_Y=q_0$ and $p_Z=1-2q_0$, and indicate the critical non-commutativity $\mathcal{I}_c$ at the transition point, as shown in FIG.~\ref{fig:fig9}. It should be pointed out that given the typical error of $q_c(L\rightarrow\infty)$ in Eq.~(\ref{eq:eq13}), the corresponding $\mathcal{I}(p_X=p_Y=q_c, p_Z=1-2q_c)$ has typical error:

\begin{equation}
    \mathcal{I}(p_X=p_Y=q_c, p_Z=1-2q_c)=0.076227\pm 0.000034
\end{equation} 
This error bar is relatively small enough to be neglected in the following analysis.

The same trend as FIG.~\ref{fig:fig4} is observed except that for odd $r$ the transition point is from area-law phase to volume-law phase whereas for even $r$ it is from area-law phase to critical phase. Remarkably, as shown in FIG.~\ref{fig:fig10}, the resulting $\mathcal{I}_c(r)$ data points for odd $r$ which are represented as black points again fall on a straight line, exhibiting the same linear dependence on the measurement range $r$ as observed for factorizable ensembles. This agreement confirms that the linear scaling of the critical non-commutativity is not specific to a particular ensemble construction, but persists across different models. Even more interestingly, the red hollow square points indicating the critical non-commutativity at the transition point for even $r$ also lie around the fitting line though the stationary phase is completely different.

\begin{figure}[t]
    \centering
    \includegraphics[width=0.9\columnwidth]{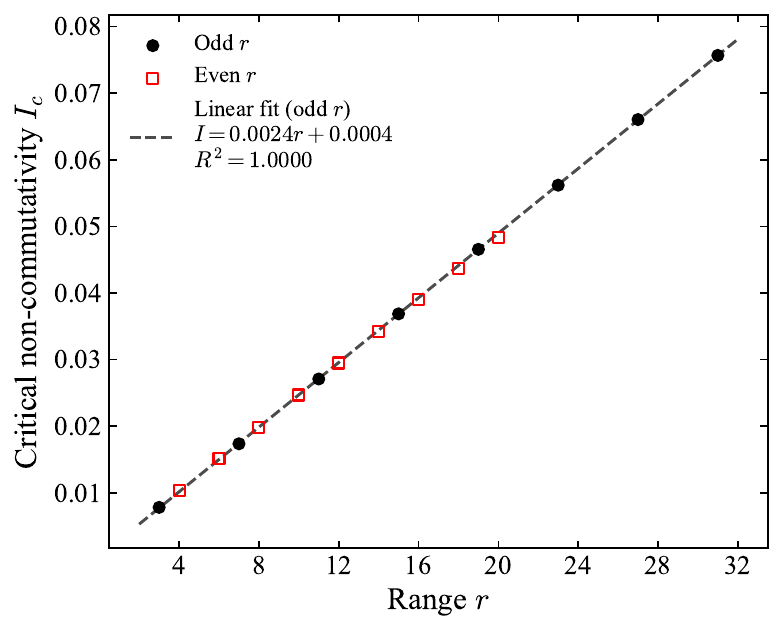}
    \caption{Critical non-commutativity $\mathcal{I}_c(r)$ as a function of the measurement range $r$ for the XYZ measurement-only model. The black dots and red hollow symbols denote values obtained by evaluating at the numerically determined transition points for odd and even $r$ respectively. The data are well described by a linear fit, $\mathcal{I}_c(r)=k_Ir+b_I$, with fitting parameters $k=0.0024, b=0.0004$ and coefficient of determination $R^2=0.99996$, demonstrating a robust linear scaling of the critical non-commutativity with the measurement range.} 
    \label{fig:fig10}
\end{figure}

In order to further testify the controlling effect of non-commutativity structures, we construct three minimal measurement ensembles: (1) $\mathcal{E}=\{XXY\}$, (2) $\mathcal{E}=\{XYXY\}$, and (3) $\mathcal{E}=\{XXXYY\}$, referred to as the cycle-3, cycle-4, and cycle-5 models, respectively. Their corresponding non-commutative structures are shown in FIG.~\ref{fig:fig11}, where the model label indicates the length of the shortest cycle. The measurement ensembles of the cycle-3 and cycle-5 models are three-partite, whereas that of the cycle-4 model is bipartite. Moreover, these three ensembles are distinguished by a finer algebraic structure. Specifically, the cycle-3 model contains a subset of measurement operators with at least three elements that are mutually anti-commuting, while the cycle-5 model does not admit any such subset of pairwise anti-commuting operators, despite being non-bipartite. Equivalently, this reflects the absence of any triangle in its associated frustration graph. We simulate the entanglement dynamics of these ensembles and plot the entanglement entropy as a function of the logarithm base 2 of subsystem size $\log_2|A|$, as shown in FIG.~\ref{fig:fig12}. The results demonstrate that the stationary states exhibit volume-law entanglement for both the cycle-3 and cycle-5 models, whereas the cycle-4 model displays a critical phase with logarithmic scaling. These results imply that a higher degree of separability in non-commutativity structure is required for a stable volume law phase. More importantly, the comparison between the cycle-3 and cycle-5 models shows that the existence of a subset of pairwise anti-commuting operators is not a necessary condition for the emergence of a volume-law phase.

\begin{figure}[t]
    \centering
    \includegraphics[width=0.98\columnwidth]{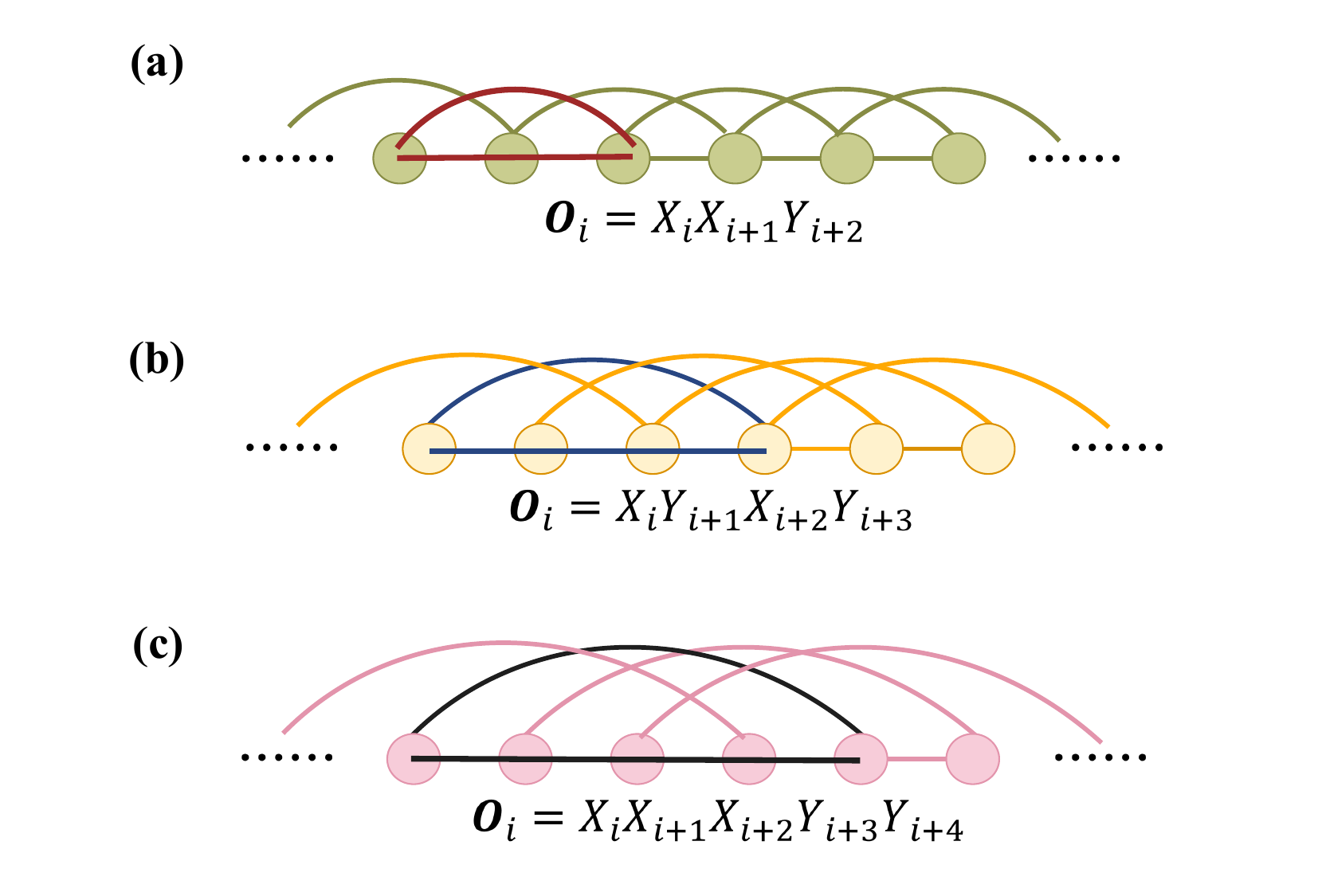}
    \caption{Non-commutative structure for the cycle-3(a), cycle-4(b), and cycle-5(c) models. The cycle-3 and cycle-5 models are non-bipartite, whereas the cycle-4 model is bipartite. Moreover, the cycle-4 and cycle-5 models do not possess a pairwise non-commuting subset structure, while such a structure exists in the cycle-3 model. The highlighted lines denote the corresponding minimal cycle structure.} 
    \label{fig:fig11}
\end{figure}
\begin{figure}[t]
    \centering
    \includegraphics[width=0.9\columnwidth]{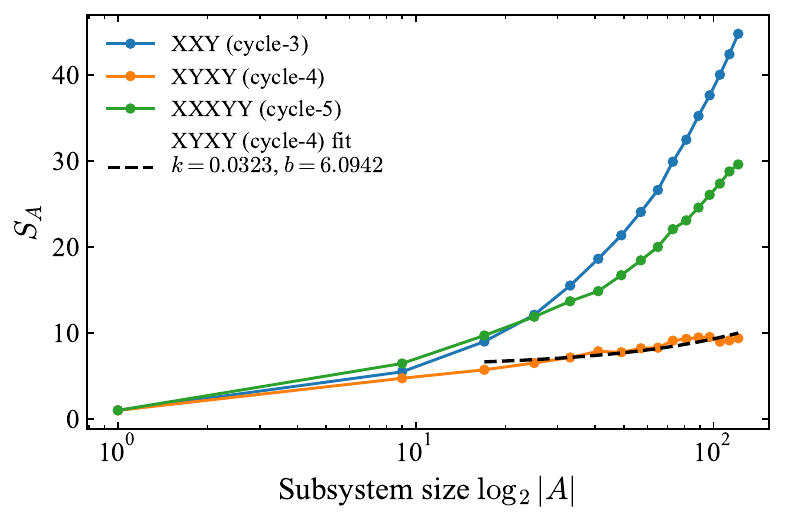}
    \caption{Stationary bipartite entanglement entropy $S_A$ versus logarithm base 2 of subsystem size $\log_2|A|$ for minimal cycle models with different frustration-graph structures: $XXY$ (cycle-3), $XYXY$ (cycle-4), and $XXXYY$ (cycle-5). The cycle-3 and cycle-5 models show volume-law scaling, whereas the cycle-4 model displays logarithmic scaling consistent with a critical phase.}
    \label{fig:fig12}
\end{figure}

\subsection{\label{sec:MRFE}Mixed-Range Factorizable Ensembles}
\begin{table}[t]
    \centering
    \begin{tabular}{lccc}
        \toprule
         & Type & Cutoff & Transition $q_0$ \\
        \midrule
        Model1 & Uniform & $r=3,4,5,6$     &  0.0938 \\
        Model2 & Uniform & $r=1,2,3,4,5,6$ &  0.1500 \\
        Model3 & Exp     & $r=3,4,5,6$     &  0.1031 \\
        Model4 & Power   & $r=3,4,5,6$     &  0.1309 \\
        \bottomrule
    \end{tabular}
    \caption{Different models adopted and their transition points along a symmetric direction in parameter space $p_X=p_Y=q_0,\ p_Z=1-2q_0$.}
    \label{tab:model_summary}
\end{table}
So far, our analysis has focused on measurement ensembles characterized by a fixed measurement range $r$. In this case, the entanglement phase transition is controlled by a critical value of the non-commutativity index $\mathcal{I}_c(r)$, which exhibits a linear dependence on $r$. An important question is whether this picture remains valid when the measurement ensemble contains Pauli strings of multiple ranges. To address this issue, we introduce mixed-range factorizable ensembles, in which the measurement range $r$ is itself a random variable drawn from a prescribed distribution $p(r)$. Physically, the measurement range plays a role analogous to the spatial extent of interactions in many-body systems, and mixed-range ensembles provide a natural setting to test the robustness of the non-commutative index beyond fixed-range models.

In our construction, each measurement operator takes the form
\begin{equation}
    O_\alpha\equiv \sigma_i\sigma_{i+1}...\sigma_{i+r-1}
\end{equation}
where the starting site $i$ is chosen uniformly along the chain, the Pauli operators $\sigma_k\in\{X,Y,Z\}$ are sampled independently with probabilities $p_X,p_Y,p_Z$, and the range $r$ is drawn from a distribution $p(r)$. The probability of selecting a given operator thus reads
\begin{equation}
    P_{O_\alpha}=\frac{1}{L}p(r)\prod_{k=i}^{i+r-1}p_{\sigma_k}
\end{equation}
with $p_X+p_Y+p_Z=1$.
We consider three representative choices of $p(r)$: a uniform distribution over a finite interval, an exponential distribution, and a power-law distribution. Besides, we choose two different ranges of $r$ for uniform distribution, as summarized in TABLE ~\ref{tab:model_summary}. This consideration originates from the fact that for fixed-range factorizable ensembles, $r=1,2$ cases do not have a volume-law phase, which implies that they may be qualitatively special. An intuitive expectation would be that the critical point of a mixed-range ensemble can be given by a probabilistic average of the fixed-range transition points,
\begin{equation}
    q_c^{naive}=\sum_r p(r)q_c(r).
\end{equation}
However, direct numerical calculations demonstrate that this naive expectation breaks down.

This failure indicates that the entanglement transition in mixed-range ensembles is not controlled by a simple average of the measurement probabilities, but rather by a more refined quantity associated with non-commutativity. Motivated by the linear relation $\mathcal{I}_c(r)=k_Ir+b_I$ observed in fixed-range ensembles, we reinterpret the slope
\begin{equation}
    k_I=\frac{\mathcal{I}_c(r)-b_I}{r}
\end{equation}
as a characteristic measure of non-commutativity per unit range for a given class of measurement ensembles. For mixed-range ensembles, we therefore define an effective slope
\begin{equation}
    k_{eff}=\sum_r p(r)\frac{\mathcal{I}_c(r)-b_I}{r}
\end{equation}
and hypothesize that the entanglement phase transition occurs when $k_{eff}$ reaches critical value. The values of $k_{eff}$ extracted at the numerically determined transition points are plotted in FIG.~\ref{fig:fig13}. In this model, we obtain the transition points by finite scale collapsing of tripartite mutual information $I_3$ \cite{doi:https://doi.org/10.1002/047174882X.ch2,PhysRevA.61.052306,PhysRevD.87.046003} which is defined by
\begin{equation}
    I_3=S_A+S_B+S_C-S_{AB}-S_{BC}-S_{CA}+S_{ABC},
\end{equation}
where $A$,$B$,$C$ are three consecutive intervals of length $L/2$. We find that, except for Model 2, all mixed-range ensembles yield values of $k_{eff}$ that are remarkably close to each other. The deviation observed in Model 2 can be traced to the inclusion of short-range measurements with $r=1,2$, which are qualitatively distinct: such measurements predominantly destroy entanglement and lack the ability to generate or reorganize long-range correlations. This observation is consistent with the absence of a volume-law phase in fixed-range ensembles with $r\le2$.

Taken together, these results suggest that the non-commutativity index remains useful beyond fixed-range ensembles when the distribution of measurement ranges is properly taken into account. 

\begin{figure}[t]
    \centering
    \includegraphics[width=0.9\columnwidth]{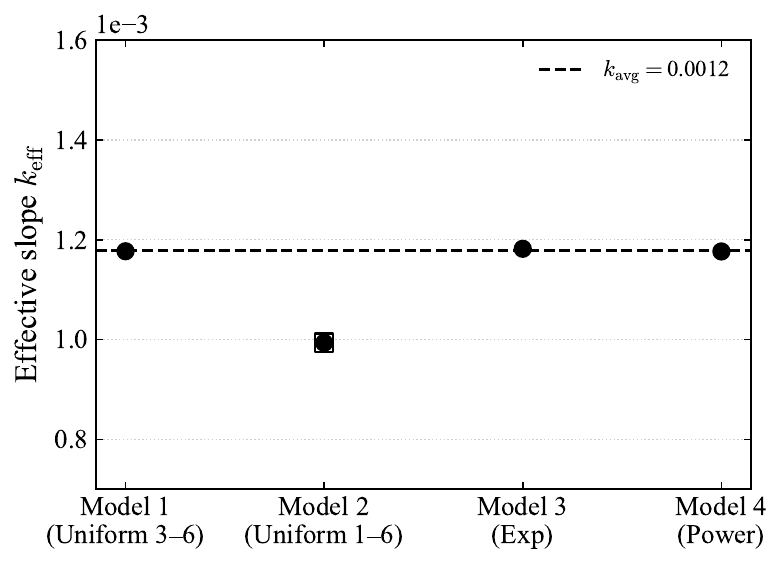}
    \caption{Effective slope $k_{eff}$ extracted at the entanglement phase transition for different mixed-range factorizable ensembles. Each symbol corresponds to a distinct choice of the range distribution $p(r)$, as summarized in Table I, with $k_{eff}$ defined by the weighted contribution of fixed-range critical non-commutativity. The horizontal dashed line marks the averaged value of $k_{eff}$ of Model 1,3,4. All models except Model 2, which includes short-range measurements with $r=1,2$, are close, demonstrating that the entanglement transition in mixed-range ensembles is controlled by the same non-commutativity criterion.}
    \label{fig:fig13}
\end{figure}

\section{\label{sec:Conclusion}Conclusion}

In this work, we have systematically investigated the entanglement phase transition in spatially and temporally homogeneous one-dimensional measurement-only circuits within stabilizer formalism, aiming to identify the role played by non-commutativity among measurement operators. Unlike unitary or hybrid circuits, where entanglement growth is driven by unitary scrambling processes, measurement-only dynamics admit no intrinsic unitary evolution. The emergence of nontrivial entanglement phases must therefore originate purely from the algebraic structure of the measurement ensemble itself.

Our main result is the identification of a quantitative ensemble-level non-commutative index, $\mathcal{I}(\mathcal{E})$, which characterizes the global strength of anti-commutation within a measurement ensemble. By employing this index to three representative measurement-only models: fixed-range factorizable ensembles, the fully correlated XYZ model, and mixed-range factorizable ensembles, we unveil the role of non-commutativity in measurement-only dynamics: For higher than bipartite ensembles, when the non-commutative index $\mathcal{I}(\mathcal{E})$ is sufficiently large, the system can be driven into a volume-law phase. The critical non-commutativity $\mathcal{I}_c$ is solely related to the measurement range $r$ and is independent of microscopic details of the measurement probabilities. The linear scaling $\mathcal{I}_c=k_Ir+b_I$ has been proven effective across different models. However, for bipartite ensembles, the volume-law phase is absent even for sufficiently large $\mathcal{I}(\mathcal{E})$. 

Before ending, it is necessary to emphasize that the simplicity of expression concerning the index referred in Eq.~(\ref{eq:eq3}) relies on two factors: 1.The stabilizer formalism, 2.The spatial and temporal homogeneity. When the measurements are beyond the stabilizer formalism, the simple commuting-or-anti-commuting algebra breaks and the entanglement quantity is measurement results relevant. Beyond stabilizer formalism, the binary function $I(M_1,M_2)$ in Eq.~(\ref{eq:eq4}) should be substituted by some measure of incompatibility \cite{Heinosaari2015Incompatibility, PhysRevLett.123.070401, RevModPhys.95.011003, Zhu_2024}. Moreover, the homogeneity of measurement ensemble directly leads to the contribution of higher order terms probably including N-body mutual incompatibility relations and time decaying factor vanishes. For inhomogeneous measurement ensembles, including the higher order terms in Eq.~(\ref{eq:eq3}) may be necessary. We will explore these points in the future work.

\section{Acknowledgments}
This work was supported by the National Natural Science Foundation of China under Grant No.61975005, the Beijing Academy of Quantum Information Science under Grants No.Y18G28, and the Fundamental Research Funds for the Central Universities.

\section*{Data Availability Statement}

The data that support the findings of this article are openly available in Ref.~\cite{HuangData2026}.

\end{document}